\mag=\magstephalf
\mag=\magstep1
\pageno=1
\input amstex
%\baselineskip = 1.0 true cm
\documentstyle{amsppt}
\TagsOnRight
\interlinepenalty=1000
%\hsize=6.5truein
%\voffset=0.3truein
%\hoffset=0.2truein
%\vsize =9.8truein
\NoRunningHeads
\pagewidth{16.5 truecm}
\pageheight{23.0 truecm}
\vcorrection{-1.0cm}
\hcorrection{-1.2cm}
\advance\vsize by -\voffset
\advance\hsize by -\voffset
%\baselineskip = 1.0 true cm
\nologo

\NoBlackBoxes
\font\twobf=cmbx12
\define \bt{{\bar t}}
\define \ii{\roman i}
\define \ee{\roman e}
\define \dd{\roman d}

\define \NLS{\roman{NLS}}

\define \qcl{{\roman{qcl}}}
%\define \qcl{{\roman{}}}
\define \reg{{\roman{reg}}}

\define \CC{{\Cal C}}

\define \XX{{\bold X}}
\define \ba{{\bold a}}
\define \tvskip{\vskip 0.5 cm}

{\centerline{\bf{Statistical Mechanics of Non-stretching 
Elastica in Three Dimensional Space}}}
%{\centerline{\bf{:On Large Polymer}}}
{\centerline{\bf{}}}

\author
\endauthor
\affil
Shigeki MATSUTANI\\
2-4-11 Sairenji, Niihama, Ehime, 792 Japan \\
\endaffil
\endtopmatter
%\baselineskip= 0.8 true cm

\subheading{Abstract}

Recently I proposed a new calculation scheme of a partition
function of an immersion object using  path integral method 
and theory of soliton (J.Phys.A (1998) {\bf 31} Mar).
I applied the scheme to problem of elastica in two-dimensional space
and Willmore surface in three dimensional space.
In this article, I will apply the scheme to elastica in
three dimensional space as a more physical model in polymer science.
Then orbit space of the nonlinear Schr\"odinger 
and complex modified  Korteweg-de Vries  equations 
can be regarded as the functional space 
of the partition function.

%\subheading{PACS numbers}:
%\baselineskip= 0.8 true cm

\document

%\newpage
%\baselineskip= 0.8 true cm

\tvskip
\centerline{\twobf \S 1. Introduction }
\tvskip

Elastica problem in two dimensional space 
$\Bbb R^2$ has long history [1,2]. 
It is known that by observing a shape of thin elastic beam,
James Bernoulli named the shape  elastica. 
It might be regarded as birth of the elastica problem and 
germination of the mathematical physics, including 
the elliptic function theory, mode analysis, nonlinear science,
elliptic differential theory, algebraic analysis
and so on.
The elastica in $\Bbb R^2$ [1,2] is defined as a  curve with
the Bernoulli-Euler functional,
$$
           E=\int \dd s\ k^2, \tag 1-1
$$
where $k$ is its curvature.

Recently I presented a new calculation scheme of a partition function
of non-stretching elasticas in $\Bbb R^2$ under the condition
 preserving its local length [3]. The partition function is formally
 defined as
$$
	Z  = \int D X \ee^{-\beta \int \dd s\ k^2}
	, \tag 1-2
$$
where $DX$ is the Feynman measure for an affine vector of a point of 
the elastica $X$ and $\beta$
is the inverse of temperature.
Goldstein and Petrich discovered that the virtual motion of 
non-stretching curve obeys the modified Korteweg-de Vries (MKdV) equation,
$$
	\partial_t k + \frac{3}{2} k^2 \partial_s k + \partial_s^3 k =0
       , \tag 1-3
$$
and  its hierarchy [5,6]. 
Using the Goldstein-Petrich scheme, I
found that the functional space of the partition function (1-2)
are completely represented by the MKdV equation (1-3).
In other words, the MKdV flows conserves the energy functional (1-1). 
The functional space (1-2) is classified by the solutions of the MKdV
equation (1-3).

After that, I applied this method to the Willmore surface in
three dimensional space $\Bbb R^3$ [4]. Instead of the MKdV equation,
there appears the modified Novikov-Veselov equation which 
classifies the functional space of the partition function.

In this article, I will investigate a partition function of
an elastica in $\Bbb R^3$ with the energy functional
$$
           E=\int \dd s\ |\kappa|^2, \tag 1-4
$$
where 
$\kappa$ is a complex curvature of the elastica in $\Bbb R^3$.
I will also require that the elastica does not stretch.

Then the partition function of an elastica in $\Bbb R^3$ 
with the energy (1-4) can be also evaluated.
Due to the non-stretching condition,
instead of  Goldstein-Petrich scheme of the 
MKdV hierarchy [3,5,6],   the Langer-Perline scheme of the
nonlinear Schr\"odinger (NLS) hierarchy and the 
complex MKdV (CMKdV)   hierarchy  appears in the calculation of the 
partition function [7,8].

Whereas the NLS equation is well known as the integrable equation
and investigated well, the properties of the CMKdV equation is not
sufficiently studied.
According to the result of Mohammad and Can [10], the different version
of the CMKdV equation does not pass the Painlev\'e test [10]. 
In this article, I will also argue the properties of the CMKdV equation
and the relation between the CMKdV and the NLS equations.

On the other hand, the study of elastic chain model of 
a large polymer is current [11]. According to recent review of a
large polymer [11], statistical mechanics of a polymer model
is closely connected with the mathematical science.
Due to the complexity, investigation of its properties  is
not simple in general.
However  it sometimes can be exactly performed owing to
deep symmetry [11]. 
In fact an exact partition function of elastic chain  with  the 
energy functional (1-4)
 was obtained by Sait\^o {\it et al.} using the path integral [12]. 
However they paid no attention upon 
isometry condition as thermal fluctuation of the path integration
even though they required isometry condition after all computations;
they summed  allover configuration space without isometry condition
rather than over restricted functional space.
It should be noted that the constraint does not commute with  
such evaluation of the partition function in general.

Thus as another limit, it is of interest to investigate the partition 
function with the energy (1-4) under the isometry condition.
One of purposes of this article is to
 investigate the partition function of a non-stretching space curve
with the energy functional (1-4) as a polymer model.

Furthermore, a space curve in $\Bbb R^3$ also interests us from
the viewpoint of the string theory [15].
Grinevich and Schmidt investigated closed condition 
of a space curve obeying the NLS equation
because a kind of its complexfication becomes a surface with 
K\"ahler metric [14]. Thus the problem is associated with the 
string theory [15].
(However as I mentioned in ref.[3], it should be noted that
the elastica absolutely differs from a string in the 
string theory, even though it influences the theory [15].)
Thus although it is not main purpose,
another hidden purpose of this article is to investigate the moduli of
non-stretching curve in $\Bbb R^3$ 
by taking into the consideration of such 
relation as a generalization to the surface problem [4,14].

The organization of this article is as follows.
 In \S 2, I will evaluate the partition function
of non-stretching elastica in $\Bbb R^3$. 
Section 3 gives a discussion of the results.

\tvskip
\centerline{\twobf \S 2. Partition Function of Non-stretching Elastica 
in $\Bbb R^3$}
\tvskip

I will denote by $\CC$ a shape of an elastica 
(a real one-dimensional curve) immersed 
in three dimensional space $\Bbb R^3$
and by $\XX(s)=(X^1,X^2,X^3)$ its affine vector, 
$$
           S^1 \ni s \mapsto X(s) \in \CC \subset \Bbb R^3  ,
           \quad \partial_s^{n}\XX(s+L)=\partial_s^{n}\XX(s), 
           \quad (n\in \Bbb N+\{0\})
           \tag 2-1
$$
where $L$ is the length of the elastica $s$ is a parameter of the curve
and $\Bbb N$ is natural number.
I consider a   closed polymer in $\Bbb R^3$; its center axis is 
a space curve $C$.
Here I will fix  the metric of the curve $\CC$ induced from 
 the natural metric of $\Bbb R^3 $; 
 $$
 \dd s = \sqrt{\dd \XX \dd \XX}. \tag 2-2
$$
As I stated in ref.[3], a reader should not confuse an elastica with
a "string" in a string theory; they are absolutely different.

There is the orthonormal system along $C$,
$({\bold n}_0, {\bold n}_1,
 {\bold n}_2)$ with fixing 
${\bold n}_0$ as the tangent unit vector; $\bold n_0 = \partial_s \XX$,
where $\partial_s:=\partial/\partial s$.
We make them, first,  satisfy the Frenet-Serret relation 
[16],
$$
    \partial_s
    \pmatrix {\bold n}_0 \\ {\bold n}_1 \\ {\bold n}_2
         \endpmatrix
    =
    \pmatrix 0   & k &  0    \\ 
             -k &  0  &  \tau   \\
               0  & -\tau &   0   \\
    \endpmatrix
    \pmatrix {\bold n}_0 \\ {\bold n}_1 \\ {\bold n}_2
\endpmatrix  .
                    \tag  2-3
$$
Here $k$ is the  curvature, 
$\tau$ is the Frenet-Serret torsion 
and they are functions of only $s$.
We rotate the orthonormal frame SO(2) fixing ${\bold a}_0:={\bold n}_0$
so that we obtain $({\bold a}_0,{\bold a}_1 ,{\bold a}_2)$
[17-19],
$$
    \partial_s
    \pmatrix {\bold a}_0 \\ {\bold a}_1 \\ {\bold a}_2 \endpmatrix
    =
    \pmatrix 0   & \kappa_1&\kappa_2 \\ 
       -\kappa_1 &  0        &  0    \\
       -\kappa_2 &  0        &   0    \\
    \endpmatrix
    \pmatrix {\bold a}_0 \\ {\bold a}_1 \\ {\bold a}_2 
                \endpmatrix  .
                    \tag 2-4
$$
where $\kappa_1:= k \cos \theta$, $\kappa_2:= k \sin \theta$
and
$$
        \theta (s):= \int^{s}_{s_0} \tau(s') \dd s'  .  \tag 2-5
$$
For convenience, we introduce a complex curvature as
$$
    \kappa :=\kappa_1+\ii \kappa_2 =
        k \ee^{\ii \theta}
              .  \tag 2-6
$$
 
In this article,
I will deal with  a non-stretching elastica in $\Bbb R^3$
with the energy functional 
$$
       E=\int_0^L \dd s\  |\kappa|^2 , \tag 2-7
$$
 which I will also  call  Bernoulli-Euler functional [3].
 
It is worth while noting  that in general, there  appear other potential 
terms in the energy functional for a general elastic rod. 
For example, there might
appear elastic torsion term,  stretching term and so on. 
An elastica is usually defined as a curve realized as a stationary point
of an energy functional related to an elastic rod, at least, 
in the meaning of the classical mechanics.
Hence the word "elastica" sometimes has ambiguity.
Depending upon the potential term, its shape might belong to 
individual class. 
Thus reader should not confuse the word "elastica" with another
one in another context.
In this article, the word of "elastica" is meaning of a curve
with the Bernoulli-Euler functional (2-7).

The elastica I deal with here is a  model of  a polymer
which can freely rotate around its center axis but does not stretching 
and is forced by the potential (2-7). In other words,
I assume that the force from the elastic torsion can be negligible
but stretching can not.
Furthermore, I will neglect the kinetic term of the elastica.
Physically speaking, I will consider the polymers 
 in the liquid whose temperature is determined and viscosity 
 is very large. I also suppose that each polymer behaves
independently and interaction among them are neglected.

Let the elastica closed and preserve 
its local infinitesimal length for even thermal fluctuation;
 it does not stretch.
Under the conditions, I will consider this partition function of 
the elastica given as [3],
$$
      \Cal Z=\int D \XX \exp
       \left(-\beta \int^L_0 \dd s\  |\kappa|^2 \right)  .
           \tag 2-8
$$
Following the calculation scheme which I proposed in refs.[3,4], 
I will evaluate
the partition function (2-8) under the non-stretching condition. 

However there is trivial affine symmetry of the centroid and direction of 
the elastica and  the partition function naturally diverges [3].
For an affine transformation (translation and rotation $g\in$S0(3)), 
$ \XX(s) \to \XX_0 + g\XX(s)$,
($\XX_0$ and $g$ are constants of $s$), 
the curvature $\kappa$ and the Bernoulli-Euler functional (2-7)
does not change;
 this is a gauge freedom and the energy functional (2-7) has
infinitely degenerate states. In the path integral method, 
I must sum over all possible states, $\Cal Z$
includes the integration over $\Bbb R^3$ 
and naturally diverges.
As well as the arguments in refs.[3,4], I will regularize it,
$$
            \Cal Z_{\reg}=\frac{\Cal Z}{\text{Vol}(\text{Aff})},
             \tag 2-9
$$
where $\text{Vol}(\text{Aff})$ is the volume of the
space related to  the affine transformation. 
By this regularization, I can concentrate
the classification of  shapes of elastica.

Next I will investigate the condition preserving local length 
even for the thermal fluctuation.
I will expand the affine vector around the point which is
an extremum point of the Bernoulli-Euler functional (2-7). I will
call the point quasi-classical point according to 
 the semi-classical method in path integral [3].
In the path integral, I must pay attention to the 
higher perturbations of $\epsilon$ 
in order to obtain an exact result.
Hence I will assume that $\XX$ is parameterized by a parameter $t$.
I will express a
perturbed affine vector $\XX$ around an extremum point $\XX_\qcl$
in the partition function (2-9) as [3,4,7,8],
$$
     \XX(s,t):=\ee^{\epsilon \partial_{t}}\XX_{\qcl}(s,t),
       \quad \epsilon \partial_{t} \XX_\qcl  =\XX_\qcl - \XX 
       + \Cal O(\epsilon^2). \tag 2-10
$$
with the relation
$$
   \partial_{t} \XX_{\qcl} = u_0 \ba_0 + u_1 \ba_1+ u_2 \ba_2,
   \quad u_a(L)=u_a(0), \quad (a=0,1,2),
     \tag 2-11
$$
where $u$'s are real function of $s$ and $t$. I will regard 
(2-11) as virtual dynamics of the curve describing the thermal 
fluctuation [3].
As  in refs.[3,7,8], due to the isometry 
condition, I require $ [\partial_{t},\partial_s]=0$ for $\XX$. 
Since $\dd s_\qcl :=\sqrt{ \partial_s 
\XX_\qcl\partial_s  \XX_\qcl}\dd s$,
 the isometry condition exactly
preserves, $\dd s\equiv \dd s_\qcl$.
Here I will note that the deformation (2-10) generally contains 
non-trivial ones
 through $u_a(s)$ and the "equation of motion" (2-11).

Let us compute the non-stretching condition 
$[\partial_{t},\partial_s]\XX_\qcl=0$.
I will introduce "velocities" 
$(\partial_t \phi_1,\partial_t \phi_2)$ as 
$$
\partial_t \ba_0\equiv \partial_{t} \partial_{s} \XX_\qcl
:=(\partial_t \phi_2 ) \ba_1-(\partial_t \phi_1 ) \ba_2 .
      \tag 2-12
$$
and 
$$
  \partial_{s}\partial_{t} \XX_\qcl 
  = (\partial_s u_0-\kappa_1 u_1 -\kappa_2 u_2) \ba_0
   +(\partial_s u_1+\kappa_1 u_0 ) 
   \ba_1+(\partial_s u_2+\kappa_2 u_0 ) \ba_2.
    \tag 2-13
$$
From the condition,
I have the relation between $\partial_t \phi_c$ 
($\phi_c := \phi_1+\ii \phi_2$) and 
a complex "velocity", $u_c := u_1 + \ii u_2$,
$$
	\partial_t \phi_c = \ii( \kappa_{\qcl} u_0+ \partial_s u_c) 
	=\ii (\kappa_{\qcl} \partial_s^{-1}
	Re((\overline{\kappa_{\qcl}} u_c)
	+\partial_s u_c)=:Q(u_c). \tag 2-14
$$ 
Here I use the notation $\kappa_{\qcl}:=\kappa_1+\ii \kappa_2$
and I introduce the pseudo-differential operator $\partial_s$ 
in the meaning of
$$
 \split
	\partial_s u_0 &= Re(\overline{ \kappa_{\qcl} }u_c)=
	(\kappa_{\qcl}\overline{ u_c}+\overline{ \kappa_{\qcl}}u_c)/2,\\
	u_0 &= \partial_s^{-1}Re(\overline{\kappa_{\qcl}} u_c)
	= \int^s \dd s' Re(\overline{\kappa_{\qcl}(s')} u_c(s')). 
 \endsplit \tag 2-15
$$
In order to find the connection between $\phi_c$ and $\kappa$,
I will also investigate the fluctuation of $\ba_a$ $(a=1,2)$.
By the virtual dynamics of $\ba_0$, differentiation of $\ba_a$ $(a=1,2)$ 
by $t$ must have the form,
$$
\partial_t \ba_1=-\partial_t\phi_2 \ba_0 - v \ba_2, \quad
\partial_t \ba_2=\partial_t\phi_1 \ba_0 +v \ba_2,
                \tag 2-16
$$
where $v$ means the rotation in the plane spanned by $\ba_a$ $(a=1,2)$.
By requirement of the isometry, the virtual dynamics
of $\ba_a$ is constrained as 
$ [\partial_{t},\partial_s]\ba_a=0$ ($a=1,2$), 
$$
   - \partial_{s}\partial_{t} \ba_1 
   = (\partial_s\partial_t \phi_2-\kappa_2 v) \ba_0
  +(\partial_t \phi_2\kappa_1 ) \ba_1
  +(\partial_t \phi_2\kappa_1+\partial_s v ) \ba_2,   
$$
$$
   - \partial_{s}\partial_{t} \ba_2 
   = -(\partial_s \partial_t\phi_1-\kappa_2 v) \ba_0
  -(\partial_t \phi_1\kappa_1 ) 
  \ba_1-(\partial_t \phi_1\kappa_2+\partial_s v ) \ba_2,
$$
$$
  -\partial_{t} \partial_{s} \ba_1 
  =\partial_t \kappa_1\ba_0 +(\kappa_1\partial_t \phi_2) \ba_1
  -(\kappa_1 \partial_t \phi_2 ) \ba_2,
$$
$$
  -\partial_{t} \partial_{s} \ba_2 =\partial_t \kappa_2\ba_0 
  +(\kappa_2\partial_t \phi_2) \ba_1
  -(\kappa_2 \partial_t \phi_2 ) \ba_2.
              \tag 2-17
$$
Hence I have the relation [7,8],
$$
	\partial_t \kappa_{\qcl} =- Q(\partial_t \phi). \tag 2-18
$$ 
Accordingly I have the relation between $\partial_t \kappa$ and 
complex velocity $u_c$ as the "equation of motion" of 
the deformation satisfied with the isometry condition [7,8],
$$
	\partial_t \kappa_{\qcl} = -Q^2 (u_c). \tag 2-19
$$
I will remark that $Q^2$ is known 
as the recursion operator of the NLS and CMKdV equations.

For this non-stretching deformation, the Bernoulli-Euler functional (2-7)
changes as
$$
    \split 
  \int |\kappa|^2 \dd s &= \int\bigl(|\kappa_{\qcl}|^2 
  + \epsilon ( \overline{ \kappa_{\qcl}} \partial_t \kappa_{\qcl}
  + \kappa_{\qcl} \partial_t \overline{\kappa_{\qcl}}) \\
   & \qquad \qquad +\epsilon^2(
  (|\partial_t\kappa_{\qcl}|^2+\overline{\kappa_{\qcl}}
  \partial_t^2 \kappa_{\qcl} 
         +\kappa_{\qcl}\partial_t^2 \overline{\kappa_{\qcl}})+
       \cdots  \bigr)\dd s \\
   &=\int\bigl(\kappa_{\qcl}^2 -
   \epsilon (\overline{ \kappa_{\qcl}} Q^2(u_0)
   + \kappa_{\qcl} \overline{Q^2(u_0)})\\
   & \qquad \qquad +\epsilon^2(
  (|\partial_t\kappa_{\qcl}|^2
  +\overline{\kappa_{\qcl}}\partial_t^2 \kappa_{\qcl} 
         +\kappa_\qcl\partial_t^2 \overline{\kappa_{\qcl}})+
       \cdots \bigr) \dd s \\
       &=:E_{\qcl}+\delta^{(1)} E_{\qcl}+\delta^{(2)} E_{\qcl}+\cdots
        .\endsplit \tag 2-20
$$

Since I wish to expand the complex curvature $\kappa$ around 
the extremum point in the functional space, 
I will require the extremum condition [3],
$$
	\delta^{(1)} E_{\qcl}=0 . \tag 2-21
$$

In this method, I will sum the weight function
over all extremum points. Since they are extremum rather than stationary
points, they need not be realized in zero temperature.

Noting the relation $\partial_s u_0  
=(\overline{\kappa_{\qcl} }u_c+\kappa_{\qcl} \overline{ u_c})/2$
 and above notices, supposed that 
$\overline{\kappa_{\qcl}}Q^2(u_c)+\kappa_{\qcl}\overline{Q^2(u_c)}$ 
could be regarded as another function 
$\overline{\kappa_{\qcl}}u_c'+\kappa_{\qcl}\overline{ u_c'}$ of the
variation of the normal direction in (2-15), I might find the relation
$$
	\int \dd s Re(\overline{ \kappa_{\qcl}} Q^2(u_c)) \sim 
	\int \dd s Re(\overline{ \kappa_{\qcl}} u_c')  
	= \int \dd s (\partial_s   u_0') =0
		.\tag 2-22
$$

I supposed that the deformation is described by one parameter $t$.
However there is no requirement that 
I must go along with only one parameter $t$
to characterize this system. 
In the calculation of the 
partition function, one must sum up the weight function
over  events if the possibility of occurrence of the events can be
considerable. 
I will search for all possible extremum points.

Furthermore in a microcanonical system at  energy
$E_0$,  the entropy $S$ of the system is defined 
as $S:=  \log Z|_{E=E_0}$ and can be regarded as
the logarithm  of the volume of the functional space.
From primitive consideration, the dimension of 
the functional space in the statistical physics
is related to the degrees of freedom corresponding to $E_0$
and the degrees of freedom of the elastica are not finite and
its dimension need not one. 

Along the line of the arguments of ref.[3], I will give up to
express the thermal fluctuation using only one parameter $t$
and I will  introduce the sequence for mathematical times $\frak t$ 
$:=(\frak t_1,\frak t_3,\frak t_5,$ $\cdots,\frak t_{2n+1},\cdots )$ 
in this system so that (2-22) is satisfied.  
I will redefine the fluctuation (2-10) and introduce infinite parameters 
family, which can sometimes become finite set as I will show later,
$$
       \split
	 \XX_{\delta \frak t} &= \ee^{(1/\sqrt{\beta} )
	 \sum_{n=0}\delta \frak t_{2n+1} \partial_{\frak t_{2n+1}}} 
	 \XX_{\qcl}\\
	 &=\XX_{\qcl} + (1/\sqrt{\beta} )
	 \sum_{n=0} \delta \frak t_{2n+1} \partial_{\frak t_{2n+1}}
	  X_{\qcl} 
	 +\Cal O(1/\beta),
	 \endsplit \tag 2-23
$$
where  $\epsilon$  was replaced with 
$(1/\sqrt{\beta} )\delta \frak t_{2n+1}$
and $\partial_{\frak t_{2n+1}} X_{\qcl}$ is expressed as
$$
	\partial_{\frak t_{2n+1}} \XX_{\epsilon}
	= u_0^{(n)} \ba_0 + u_1^{(n)} \ba_1+ u_2^{(n)} \ba_2,
	\quad
        u_0^{(n)} = \partial_s^{-1} Re(\bar \kappa_{\qcl} u_c^{(n)}), 
        \quad
        u_c^{(n)} =  Q^{2n}(u_c^{(0)}) .
        \tag 2-24
$$
The virtual equations of motion for the deformation are expressed as
$$
	\partial_{\frak t_{2n+1}} \kappa = Q^{2n}(u_c^{(0)}). 
	\tag 2-25
$$
Thus (2-25) represents the thermal fluctuation which conserves the local
length.

However it should be noted that  
there are two manifest symmetries in this system;
one exhibits the symmetry of choice of the origin $s$
and another is for the symmetry of U(1) phase of 
$\kappa$;
the later one is the same as the choice of the $s_0$ at
the integration (2-5).
For the transformation $\kappa(s) \to \ee^{\ii t}\kappa(s-\bt)$,
the partition function is invariant.

I require that the virtual motions must include such manifest symmetries 
$$
	\partial_{\bt_1} \kappa_{\qcl}=\partial_s \kappa_{\qcl}
           \tag 2-26
$$
and
$$
	\partial_{t_1} \kappa_{\qcl}=\ii \kappa_{\qcl}. 
	 \tag 2-27
$$

As in refs.[3,4], instead of the single deformation parameter, 
I will assign the infinite dimensional parameters in (2-23)
to those which  fulfill this requirement;
$\frak t:=(\bold t,\bar{\bold t})=(t_1,t_3,\cdots,\bt_1,\bt_3,\cdots)$.
In terms of these, I will investigate the moduli space of 
the partition function (2-9).
In other words 
I will give a minimal set of the virtual equations of motion, 
which is satisfied with this physical  requirement
that the deformation contains the manifest symmetries (2-26) and (2-27),
$$
       \partial_{\bt_{2n+1}} \kappa_{\qcl} 
       =(- Q^2)^{n}( \partial_s \kappa_{\qcl}) , \quad 
      \partial_{\bt_{2n+1}} \kappa_{\qcl}
      =-Q^{2} (\partial_{\bt_{2n-1}} \kappa_{\qcl}), \quad
           (n=1,2,\cdots).
           \tag 2-28
$$
$$
       \partial_{t_{2n+1}} \kappa_{\qcl} 
       =(- Q^2)^n( \ii \kappa_{\qcl}) , \quad 
      \partial_{t_{2n+1}} \kappa_{\qcl}
      =-Q^{2} (\partial_{t_{2n-1}} \kappa_{\qcl}), \quad
           (n=1,2,\cdots).
           \tag 2-29
$$
They are  the CMKdV  and the NLS hierarchies respectively. 

As stated in the introduction, the properties of the 
CMKdV equation is not well-known as far as I know.
It has not been concluded that it is soliton equation yet. 
However even though it might not be integrable,
properties of the CMKdV hierarchy and the CMKdV equation 
 are very regular as I show as follows.

As the NLS hierarchy, a solution of the $n$-th CMKdV equation,
$$
	\partial_{\bt_{2n+1}}\kappa_{\qcl}-
	(-Q^2)^n(\partial_s\kappa_{\qcl})=0, \tag 2-30
$$
is satisfied with the simultaneous equations
by introducing unknown parameter $\bt_{2n-1}$,
$$
\partial_{\bt_{2n+1}} \kappa_{\qcl}
      -Q^{2} (\partial_{\bt_{2n-1}} \kappa_{\qcl})=0,
      \quad
    \partial_{\bt_{2n-1}}\kappa_{\qcl}
    =(-Q^2)^{n-1}(\partial_s\kappa_{\qcl}) . \tag 2-31
$$
This is a kind of B\"acklund transformation.
Thus by ladder calculations, it can be proved that
the solutions of the  higher order equations 
belonging to the CMKdV hierarchy  
are also satisfied with  the CMKdV equation,
$$
     \partial_{\bt} \kappa + \frac{3}{2} |\kappa|^2 \partial_s \kappa 
     + \partial_s^3 \kappa=0.
     \tag 2-32
$$
In other words, the nontrivial deformation obeys the CMKdV  equation
as the soliton hierarchy does.
(One might have a question why the ladder relation terminates at $\bt=\bt_3$
rather than $\bt=\bt_1$. From (2-26) $\bt_1$ is determined as 
$\bt_1 \equiv s + s_0$ and thus $\bt_1$ is not an unknown parameter in
the sense of (2-31). Thus (2-32) is a minimal non-trivial equation.)
Further it is worth while noting that for intrinsically real initial condition
 
$\kappa\in \ee^{\ii \alpha_0} \Bbb R$ for all $s$ and constant $\alpha_0$ for
$s$ and $t$, the CMKdV  equation is reduced to  the MKdV equation.
Thus the solution space of the CMKdV equation has the completely
integrable region in the meaning of the soliton theory.

Similarly,
I have the NLS equation as nontrivial deformation of the NLS
hierarchy,
$$
 \ii \partial_{t} \kappa + \frac{1}{2} |\kappa|^2  \kappa 
     + \partial_s^2 \kappa=0.
    \tag 2-33
$$
For the NLS equation, this reduction can be naturally justified 
in  the Jacobi variety of the hyperelliptic curve as a solution 
space [20].

Furthermore it is a very remarkable fact that for the variation 
of $\bt$ obeying the CMKdV  equation,
the Bernoulli-Euler functional (2-7) is invariant,
 $$
  -\partial_\bt  \int \dd s\ |\kappa(s,t,\bt)|^2
  =\int \dd s 
  \partial_s ( \frac{3}{4} |\kappa|^4 
  +(\partial_s^2 \bar \kappa) \kappa +
  \bar \kappa\partial_s^2  \kappa
  -|\partial_s \kappa|^2)
  =0, \tag 2-34
$$
as the NLS flows conserves the first integral,
$$
  \partial_t  \int \dd s\ |\kappa(s,t,\bt)|^2
  =-\ii \int \dd s \partial_s 
  ((\partial_s \bar \kappa) \kappa-\bar \kappa\partial_s  \kappa)
  =0. \tag 2-35
$$
Thus regardless of integrability of the CMKdV equation,
it was clarified that its properties are very regular.

Here I will comment upon the result of Mohammad and Can [10]. 
They investigated the "complex MKdV" equation and concluded
that it is not a soliton equation. However their "complex
MKdV equation" is expressed as
$$
     \partial_{\bt} \kappa + 
     \frac{1}{2} \partial_s(|\kappa|^2  \kappa )
     + \partial_s^3 \kappa=0,
     \tag 2-36
$$
which is a kind of "complexification" of the MKdV equation but
differs from (2-32).
Thus their result does not directly affect the studies on
the integrability of our CMKdV equation (2-32).

As I obtained the isometry deformation which includes the manifest
symmetries (2-26) and (2-27),  I will consider, here, the partition 
function (2-9).

Since the CMKdV and NLS problems  are initial value problems, 
for any regular shape of elastica
satisfied with the boundary conditions, the "time" $t$ and $\bt$
developments of the curvature are uniquely determined . 
Furthermore noting that if one gives the real value $\kappa$,
 $\kappa$ goes on real in the "time" $\bt$ development 
 of the CMKdV equation (2-32)
whereas for the NLS equation (2-33) its "time" $t$ 
development includes 
the complex value due to the pure imaginary in 
the first term in (2-33).
Thus the "time" $\dd t$ and $\dd \bt$ are expected 
orthogonal in the moduli of the CMKdV and NLS equations.
The "time" developments of  both equations differ each other.
In other words, for a given  regular curve,
there exist individual families of the solutions of the CMKdV (2-32)
and NLS (2-33) equations
 which contain the given curve as an initial condition. 
Due to relations (2-34) and (2-35),
during the motion of $t$  and $\bt$, the Bernoulli-Euler functional (2-7)
does not change its value.
Hence  the deformation parameter $t$ and $\bt$ draw the trajectories of
the functional space which have the same  value of the 
Bernoulli-Euler functional (2-7).

In the case that I immersed an elastica in $\Bbb R^2$, the 
thermal fluctuation obeys the MKdV equation and there 
appears only one sort of hierarchy or the MKdV hierarchy [3].
In this article, the codimension of the immersion of the elastica   
in $\Bbb R^3$ is two while the former problem is one [3].
Accordingly it is natural that 
there appear twice degrees of freedom of the elastica in $\Bbb R^2$,
$t$ and $\bt$ for the elastica in $\Bbb R^3$.

Thus I can formally estimate the functional space for each 
functional value.
By  the "time" development of $t_3$ and $\bt_3$, I can classify the
functional space of the partition function (2-9).
In other words by investigating the moduli of the CMKdV and NLS equations
which are satisfied with the boundary conditions,
$$
       \kappa(0)=\kappa(L), \quad   \XX_\qcl(0)=\XX_\qcl(L), \tag 2-37
$$
the measure of the functional integral $\dd \mu$ can be decomposed as,
$$
            \dd \mu= \sum_E \dd \mu_E. \tag 2-38
$$
So I denote by  $\Xi_E$  the set of these trajectories which occupy 
the same energy $E$.

Hence the partition function can be represented as
$$
           \Cal Z_\reg = \int \dd \mu \exp(-\beta E )
           =\sum_E \exp(-\beta E) \int_{\Xi_E} 
           \dd \mu_E =\sum_E \exp(-\beta E) \text{Vol}(\Xi_E)
                     , \tag 2-39
$$
where 
$$
           \text{Vol}(\Xi_E)= \int_{\Xi_E} \dd \mu_E \tag 2-40
$$
is the volume of the trajectories $\Xi_E$.

It is known that
any solutions of the  NLS equation (2-33) can be also expressed by the 
hyperelliptic function and its modulus agrees with the
modulus of the hyperelliptic curves [13,20].
Grinevich and Schmidt studied the moduli of the NLS equation (2-33)
whose corresponding space curve is satisfied with the boundary condition 
(2-37).
For the NLS equation (2-33),  there are 
 infinite Jacobi varieties who have the same energy $E$ in general. 
Thus it is expected that 
the CMKdV equation is connected among these Jacobi varieties induced
from the NLS equation (2-33).

As well as  the arguments in ref.[3], 
even though I introduced the infinite dimensional coordinates
 $ t$ in (2-23), they are  reduced to finite
  dimensional space, as the Jacobi variety of a hyperelliptic
 curve with finite dimension is embedded in the universal 
 grassmannian manifold.
Using the genus $g$ of the hyperelliptic curves, I will evaluate
the subspace and submesure of $(\Xi_E,\dd \mu_E)$ as NLS part.
The NLS part $(\Xi_E^{\NLS},\dd \mu_E^{\NLS})$ can be decomposed as
$ (\Xi_E^{\NLS},\dd \mu_E^{\NLS}) 
=\coprod_g (\Xi_E^{\NLS(g)},\dd \mu_E^{\NLS(g)})$.

For the case of a solution represented by 
the hyperelliptic function of
genus $g$ which is satisfied with (2-37), 
$\dd \mu_E^{\NLS(g)}$ is  expressed as
 $\dd t_3 \wedge \dd t_5 \wedge$
$\cdots \wedge \dd t_{2g-1}$.
For each point, there is CMKdV flows. Even though 
it has not been confirmed that 
the trajectories of the CMKdV equation are linear 
and  regarded as the vector space, 
it is clear that their cotangent space is flat and can be regarded as the 
vector space locally. Thus I can locally express the measure of $\dd \mu_E^{(g
)}$
as
$$
	\dd \mu_E^{(g)}=\dd t_3\wedge\dd t_5 \wedge 
	\cdots\wedge\dd t_{2g-1}\wedge 
	\dd \bt_3\wedge\bt_5\wedge \cdots\wedge\dd\bt_{2g-1} . 
	 \tag 2-41
$$
Here I remove $\dd t_1 \wedge \dd \bt_1$ in the measure because it 
exhibits trivial symmetries [3].
(2-41) is a subset of the infinite dimensional deformation parameters 
$\frak t$ in (2-23).  
Hence (2-40) becomes  
$$
	\text{Vol}(\Xi_E)=\sum_g \text{Vol}(\Xi_E^{(g)})=
          \sum_g \int_{\Xi_E^{(g)}} \dd \mu_E^{(g)} .  \tag 2-42
$$

By exchanging the coordinate $\dd \frak t_i$ and $\dd \frak t_j$
of multi-times $\frak t$,
 the volume of $\Xi_E^{(g)}$ is estimated by the unit of 
the elastica  length $L$.
Since the dimension of the Bernoulli-Euler functional 
$E$ is the inverse of length and 
 $\beta /$[length] is order unit,
the multiple of the length
 can be interpreted as
the multiple of the inverse temperature $\beta^{-1}$.
Hence the  sum of terms with  different dimensional 
volume which appear in (2-39) can be regarded as
expansion of power of $\beta$.

\tvskip

\centerline{\twobf \S 3. Discussion}
\tvskip

In this article, I gave a calculation scheme of the partition 
function of elasticas in $\Bbb R^3$ in terms of solutions of
 the CMKdV equation (2-32) and NLS equation (2-33).
Even though I could not give a concrete form of the partition 
function (2-9),
I showed that its formal expansion is given by (2-39).
As I thought that this scheme is based upon the soliton 
theory in refs.[3,4], I  can not deny that it might be  beyond the 
integrable system. 
In fact the CMKdV equation might be connected with 
 the deformation of the 
Jacobi variety induced by the NLS equation (2-33).
Hence I believe that this formulation might shed a new light upon the
theories of the immersed object and its quantization (or evaluation
of the partition function). 

Here I will mention the knot configuration.
Since the NLS and CMKdV equations are initial value problems, 
the solution space
includes any configurations of a space curve in $\Bbb R^3$.
In other words, they also include any knot configurations and so 
I need not pay any attention upon the ambient isotopy [21]. 
In fact the trajectories of NLS equation classify space curves immersed
in $\Bbb R^3$ rather than ones embedded in $\Bbb R^3$;
crossings are allowed and its topology disables us to 
distinguish such knot invariances or ambient isotopy.
Since the knot configuration is physically discriminated by means 
of long range force such as
the electromagnetic force and this theory in this article does not
include such force, this notion can be physically interpreted.
If one wishes to consider the knot configuration in this system,
it might be related to the gauged NLS equation [22].

Next I will give two comments on the CMKdV equation.
First, one might have a question why I need the CMKdV equation
whereas the solution space of the NLS equation 
includes any configurations of a space curve in $\Bbb R^3$.
I have been dealing with the measure of the functional space.
An uncountable set of $\Bbb R$ becomes $\Bbb R^2$ if the elements 
are measurable and one can define $\Bbb R^2$ topology in the set. 
In the similar meaning, I need CMKdV
equation in order to introduce the natural measure in the 
functional space. 

The solutions of the NLS equation are described in terms of the 
hyperelliptic functions [13,20]. A hyperelliptic curve is embedded
in a Jacobi variety. The trajectories of the NLS equation,
($t_1,t_3,\cdots,t_{2g-1}$), form the vector structure of the 
Jacobi variety. The NLS flow, which obeys the NLS equation (2-33),
covers a subset of Jacobi variety. In each Jacobi variety, there exists
compact subset as orbits of NLS flows. 
The individual Jacobi varieties are distinguished by
points in the Siegel upper half space [20].
Since the CMKdV flows are perpendicular with
the NLSE flows, 
the CMKdV flows might connect the different Jacobi
varieties of solutions of the NLS equation.
Thus I will conjecture, as the second comment upon the CMKdV 
equation, that the moduli of 
the CMKdV equation might be realized in the Siegel upper half space.
It reminds me of the facts the theta function of the elliptic 
curve obeys the heat equation over the Siegel upper half plane,
 which is not integrable in the
sense of the kinematic theory such as the soliton theory.
The integrability of the soliton theory is associated with
the time inversion symmetry and time translational symmetry,
and the solutions are acted by a (continuous) group. 
On the other hand, the solution space of the heat equation 
is acted by only semi-group and thus it is not "integrable"
in general.
By complexfication of the heat equation, imaginary time
heat equation, or Schr\"odinger equation is kinematic 
equation and integrable in the sense of the kinematic theory.
 Thus even though 
the CMKdV equation includes integrable solutions as kinematic 
region [23], I have a question regarding role of the CMKdV equation 
in Jacobi varieties of the hyperelliptic curves;
is it in the framework of the integrable system? 
However in this stage, I cannot explicitly express the role of the 
CMKdV equation because there are few studies on the CMKdV equation.
I state that the properties of the CMKdV should be investigated.

Finally I will comment upon the higher dimensional 
elastica problem, {\it e.g.}, 
 an elastica in n-dimensional space $\CC \subset \Bbb R^n$. 
The codimension of the elastica becomes $n-1$ and thus 
instead of $\frak t=(\bold t, \bar {\bold t})$,  there
appear $(n-1)$ sets of  infinite dimensional parameters
 $\frak t=(\bold t^{(1)},\bold t^{(2)},\cdots,\bold t^{(n-1)})$.
  As there appeared $U(1)$-bundle in this article,
they represent the ($n-2$)-dimensional inner sphere of 
sphere bundle over the elastica $\CC$
and the normal radius direction of $\CC$.
Thus there is naturally a principal bundle over $\CC$.
In other words, one can add the group structure over the equations. 
Thus the generalized MKdV equation naturally appears [8,24]
and it is expected that my computation scheme of the partition function 
can be extended.

%\newpage

\tvskip
\centerline{\twobf Acknowledgment}
\tvskip

I would like to thank  Y.~\^Onishi, Prof.~T.~Tokihito,
Prof.~K.~Nishinari and Prof.~S.~Saito for 
critical discussions and  continuous encouragements
and Prof.~B.~G.~Konopelchenko for helpful comment,
sending me his interesting works and encouragement.
Especially Prof.~K.~Nishinari thought me that the CMKdV equation
(2-33) has non-trivial $N$-soliton solutions. 
I am grateful Prof.~P.~Grinevich  for sending me his interesting works.
I also acknowledge that the seminars on
differential geometry, topology, knot theory and group theory with
Prof. K.~Tamano influenced this work.

%\enddocument
%\newpage

\enddocument